\documentclass{ckm}                 % twocolumn proceedings style
\usepackage{txfonts}   % if you have it, to get Times family
\confname{Workshop on the CKM Unitarity Triangle, IPPP Durham, April 2003}

\newcommand{\ba}{\!\!\!\!}

\newcommand{\nn}{\nonumber}
\newcommand{\IM}{\mbox{\rm Im}}
\newcommand{\eqn}[1]{(\ref{#1})}
\newcommand{\mev}{\mbox{\rm MeV}}
\newcommand{\gev}{\mbox{\rm GeV}}
\def\npb#1#2#3{    {\it Nucl. Phys.}~B {\bf #1}, #3 (#2)}
\def\npps#1#2#3{   {\it Nucl. Phys. Proc. Suppl.} {\bf #1}, #3 (#2)}
\def\plb#1#2#3{    {\it Phys. Lett.}~B {\bf #1}, #3 (#2)}

\def\prd#1#2#3{    {\it Phys. Rev.}~D {\bf #1}, #3 (#2)}

\def\prl#1#2#3{    {\it Phys. Rev. Lett. }{\bf #1}, #3 (#2)}

\def\zpc#1#2#3{    {\it Z. Phys.}~C {\bf #1}, #3 (#2)}

\def\epjc#1#2#3{   {\it Eur. Phys. J.}~C {\bf #1}, #3 (#2)}
\def\jhep#1#2#3{   {\it JHEP  }{\bf #1}, #3 (#2)}

\title{A novel route to $V_{us}$}

\author{Matthias Jamin}
\address{Institut f\"ur Theoretische Physik, Universit\"at Heidelberg,
 Philosophenweg 16, D-69120 Heidelberg, Germany}

\begin{document}

\begin{abstract}
Already in the past, hadronic $\tau$ decays have served as an interesting
source to obtain information on the parameters of the Standard Model, like
the strong coupling $\alpha_s$ or the strange quark mass. Below it will be
shown that the approach to obtain the strange mass from the hadronic
$\tau$-decay width can be turned into a determination of the CKM matrix element
$V_{us}$, once the strange mass is used as an input from other sources. At
present, we obtain $V_{us}=0.2179\pm 0.0045$, where the error is completely
dominated by the experimental uncertainty, and can be improved through an
improved measurement of the hadronic $\tau$-decay rate into strange particles
$R_{\tau,S}$.
\end{abstract}

\maketitle

\vskip 0.7 cm 

\section{Introduction}

Already more than a decade ago it was realised that the hadronic decay of the
$\tau$ lepton could serve as an ideal system to study low-energy QCD under
rather clean conditions \cite{bnp:92}. In the following years, detailed
investigations of the $\tau$ hadronic width as well as invariant mass
distributions have served to determine the QCD coupling $\alpha_s$ to a
precision competitive with the current world average \cite{aleph:98,opal:99}.
More recently, the experimental separation of the Cabibbo-allowed decays and
Cabibbo-suppressed modes into strange particles opened a means to also
determine the mass of the strange quark
\cite{gjpps:03,cdghpp:01,dchpp:01,km:00,kkp:00,pp:99,aleph:99,ckp:98,pp:98},
one of the fundamental QCD parameters within the Standard Model.

Until recently, strange quark mass determinations from hadronic $\tau$ decays
suffered from sizeable uncertainties due to higher order perturbative
corrections. These result from large QCD corrections to the contributions of
scalar and pseudoscalar correlation functions \cite{bnp:92,pp:98,ck:93,mal:98a}
which are additionally amplified by the particular weight functions which
appear in the $\tau$ sum rule. However, a natural remedy to circumvent this
problem is to replace the QCD expressions of scalar and pseudoscalar
correlators by corresponding phenomenological hadronic parametrisations
\cite{gjpps:03,mk:01,km:00,pp:99,aleph:99}, which turn out to be more precise
than their QCD counterparts, since the by far dominant contribution stems from
the known kaon pole.

Additional suppressed contributions to the pseudoscalar correlators come from
the pion pole as well as higher exited pseudoscalar states whose parameters
have recently been estimated \cite{mk:02}. The remaining strangeness-changing
scalar spectral function has been extracted very recently from a study of S-wave
$K\pi$ scattering \cite{jop:00,jop:01} in the framework of chiral perturbation
theory ($\chi$PT) with explicit inclusion of resonances. The resulting scalar
spectral function was then employed to directly determine the strange quark
mass from a purely scalar QCD sum rule \cite{jop:02}. On the other hand, now
we are also in a position to incorporate this contribution into the $\tau$ sum
rule.

Nevertheless, as was already realised in the first works on the strange
mass determination from the Cabibbo-suppressed $\tau$ decays, $m_s$ turns out
to depend sensitively on the element $V_{us}$ of the quark-mixing (CKM) matrix.
With the theoretical improvements in the $\tau$ sum rule mentioned above, in
fact $V_{us}$ represents one of the dominant uncertainties in the determination
of the strange mass. Thus it appears natural to turn things around and, with
an input of $m_s$ as obtained from other sources, to actually determine
$V_{us}$. It is then found that by far the dominant error on $V_{us}$ results
from the experimental uncertainty on the hadronic $\tau$-decay rate into
strange particles $R_{\tau,S}$, which should be improvable in future
experimental analyses.

\section{Theoretical framework}

The main quantity of interest for the following analysis is the hadronic decay
rate of the $\tau$ lepton,
\begin{equation}
\label{RTauex}
R_\tau \equiv \frac{\Gamma(\tau^-\to{\rm hadrons}\,\nu_\tau(\gamma))}
{\Gamma(\tau^-\to e^-\bar\nu_e\nu_\tau(\gamma))} =
R_{\tau,V} + R_{\tau,A} + R_{\tau,S} \,.
\end{equation}
Theoretically, it can be expressed as an integral of the spectral functions
$\IM\,\Pi^T(s)$ and $\IM\,\Pi^L(s)$ over the invariant mass $s=p^2$ of the
final state hadrons \cite{bnp:92},
\begin{equation}
\label{RTauth}
R_\tau = 12\pi\!\!\int\limits_0^{M_\tau^2}\!\frac{ds}{M_\tau^2}\,\biggl(
1-\frac{s}{M_\tau^2}\biggr)^2 \biggl[\biggl(1+\frac{2s}{M_\tau^2}\biggr)
\IM\Pi^T(s)+\IM\Pi^L(s)\biggr] .
\end{equation}
The appropriate combinations of two-point correlation functions which
appear in eq.~\eqn{RTauth} are given by
\begin{eqnarray}
\Pi^J(s) &\ba\equiv\ba& |V_{ud}|^2\,\Big[\,\Pi^{V,J}_{ud}(s) +
\Pi^{A,J}_{ud}(s)\,\Big] \nn \\
&\ba+\ba& |V_{us}|^2\,\Big[\,\Pi^{V,J}_{us}(s) + \Pi^{A,J}_{us}(s)\,\Big]\,,
\end{eqnarray}
with $V_{ij}$ being the corresponding matrix elements of the CKM matrix.
As has been indicated in eq.~\eqn{RTauex}, experimentally, one can disentangle
vector from axialvector contributions in the Cabibbo-allowed ($\bar ud$)
sector, whereas such a separation is problematic in the Cabibbo-suppressed
($\bar us$) sector. The superscripts in the transversal and longitudinal
components denote the corresponding angular momentum $J=1$ ($T$) and $J=0$
($L$) in the hadronic rest frame.\footnote{Further details on our notation
can be found in refs.~\cite{bnp:92,gjpps:03}.}

Additional information can be inferred from the measured invariant mass
distribution of the final state hadrons. The corresponding moments
$R_\tau^{kl}$, defined by \cite{dp:92b}
\begin{equation}
\label{Rtaukl}
R_\tau^{kl} \equiv \int\limits_0^{M_\tau^2} \, ds \,
\biggl( 1 -\frac{s}{M_\tau^2} \biggr)^k \biggl(\frac{s}{M_\tau^2}\biggr)^l\,
\frac{d R_\tau}{ds} = R_{\tau,V+A}^{kl} + R_{\tau,S}^{kl} \,,
\end{equation}
can be calculated theoretically in analogy to $R_\tau = R_\tau^{00}$. In the
framework of the operator product expansion (OPE), $R_\tau^{kl}$ can be written
as
\cite{bnp:92}:
\begin{eqnarray}
R_\tau^{kl} &\ba=\ba& 3\,S_{{\rm EW}}\biggl\{\,\Big(|V_{ud}|^2+|V_{us}|^2\Big)
\,\Big( 1 + \delta^{kl(0)} \Big) \nn \\
&\ba+\ba& \sum\limits_{D\geq2}\Big( |V_{ud}|^2\,\delta_{ud}^{kl(D)} +
|V_{us}|^2\,\delta_{us}^{kl(D)} \Big)\,\biggr\} \,.
\end{eqnarray}
The electroweak radiative correction $S_{{\rm EW}}=1.0201\pm 0.0003$
\cite{ms:88,bl:90,erl:02} has been pulled out explicitly, and $\delta^{kl(0)}$
denotes the purely perturbative dimension-zero contribution. The symbols
$\delta_{ij}^{kl(D)}$ stand for higher dimensional corrections in the OPE from
dimension $D\geq 2$ operators which contain implicit suppression factors
$1/M_\tau^D$.

The separate measurement of Cabibbo-allowed as well as Cabibbo-suppressed decay
widths of the $\tau$ lepton \cite{aleph:99} allows one to pin down the flavour
SU(3)-breaking effects, dominantly induced by the strange quark mass. Defining
the difference
\begin{equation}
\label{delRtaukl}
\delta R_\tau^{kl} \;\equiv\; \frac{R_{\tau,V+A}^{kl}}{|V_{ud}^2|} -
\frac{R_{\tau,S}^{kl}}{|V_{us}^2|} \;=\;
3\,S_{{\rm EW}}\sum\limits_{D\geq 2}\Big(\delta_{ud}^{kl(D)} -
\delta_{us}^{kl(D)}\Big) \,,
\end{equation}
many theoretical uncertainties drop out since these observables vanish in the
SU(3) limit. In particular, they are free of possible flavour-independent
instanton as well as renormalon contributions which could mimic dimension-two
corrections.

\section{Longitudinal contributions}

Before we discuss the determination of $m_s$ and $V_{us}$ from hadronic
$\tau$ decays, let us first comment on the longitudinal contributions to
eq.~\eqn{RTauth}. As was already remarked above, these contributions are
plagued with huge perturbative higher order corrections, which in previous
analyses resulted in large corresponding uncertainties for the strange quark
mass. This problem can be circumvented by replacing the theoretical
expressions with corresponding contributions resulting from phenomenological
parametrisations of the relevant spectral functions.

\begin{table}[thb]
\renewcommand{\arraystretch}{1.2}
\begin{center}
\begin{tabular}{|c|cc|}
\hline
 & Phenom: & Theory: \\
\hline
$R_{us,A}^{00,L}$ & $-0.135\pm0.003$ & $-0.144\pm0.024$ \\
$R_{us,V}^{00,L}$ & $-0.028\pm0.004$ & $-0.028\pm0.021$ \\
$R_{ud,A}^{00,L}$ & $-(7.77\pm0.08)\cdot 10^{-3}$ &
                    $-(7.79\pm0.14)\cdot 10^{-3}$ \\
\hline
\end{tabular}
\end{center}
\caption{Comparison of theoretical and phenomenological longitudinal
contributions to the (0,0) moment of the $\tau$ sum rule.
\label{tab1}}
\end{table}

A numerical comparison of theoretical and phenomenological longitudinal
contributions $R_{ij,V/A}^{kl,L}$ for the (0,0) moment is presented in
table~\ref{tab1}.\footnote{For the precise definition of the $R_{ij,V/A}^{kl,L}$
see ref.~\cite{gjpps:03}} On the phenomenological side, the dominant
axialvector $(us)$ contribution originates from the strange pseudoscalar
mesons, by far the largest part coming from the kaon pole. Also the next
two higher exited resonances have been included with a standard Breit-Wigner
resonance shape and resonance parameters as recently estimated in \cite{mk:02}.
Similarly, the axialvector $(ud)$ contribution is due to the pion pole as
well as higher exited states. However, because of the smaller pion mass, its
contribution is much suppressed. The dominant uncertainties in the
phenomenological results in table~\ref{tab1} are due to the errors in the
decay constants $f_K$ and $f_\pi$, as well as the higher resonance decay
constants.

The longitudinal vector $(us)$ contribution is related to the corresponding
scalar spectral function which has been calculated in ref.~\cite{jop:02}.
Below $2\,\gev$ the dominant hadronic systems which contribute in this channel
are the $K\pi$ and $K\eta'$ states with $K_0^*(1430)$ being the lowest lying
scalar resonance. The scalar $us$ spectral function can then be parametrised
in terms of the scalar $F_{K\pi}(s)$ and $F_{K\eta'}(s)$ form factors. These
form factors were obtained in \cite{jop:01} from a coupled-channel
dispersion-relation analysis. The S-wave $K\pi$ scattering amplitudes which
are required in the dispersion relations were available from a description
of S-wave $K\pi$ scattering data in the framework of unitarised $\chi$PT with
resonances \cite{jop:00}.

Generally, from table~\ref{tab1} one observes that phenomenological and
theoretical values for the $R_{ij,V/A}^{00,L}$ are in rather good agreement.
Nevertheless, because of the large uncertainties from higher order perturbative
corrections in the theoretical expressions, the phenomenological values are
more precise and thus in the following they will be employed for the analysis
of the hadronic $\tau$ decays.

\section{Strange quark mass}

The leading term in the SU(3)-breaking difference of eq.~\eqn{delRtaukl} is
due to the dimension-2 contribution proportional to $m_s^2$. Thus it appears
natural to determine the strange quark mass by comparing experimental and
theoretical results for $\delta R_\tau^{kl}$. In practice, the moments (0,0)
to (4,0) have been utilised in the phenomenological analysis. For low $k$, the
higher-energy region of the experimental spectrum, which is less well known,
plays a larger role and thus in this region the experimental uncertainties
dominate the strange mass determination, whereas for higher $k$ more emphasis
is put on the lower-energy region, and there the theoretical uncertainties
dominate.

In addition, as will be exploited in more detail in the next section, the
strange mass extracted from the $\tau$ sum rules also displays a sizeable
dependence on $V_{us}$, which is strongest for $k=0$ and becomes weaker for
larger $k$. The smallest combined uncertainty is then found for the (3,0)
moment which with present experimental and theoretical errors represents an
optimal choice. Taking a weighted average of the strange mass values obtained
for the different moments we then find
\begin{equation}
\label{msLR}
m_s(2\,\gev) = 103 \pm 17 \, \mev \,,
\end{equation}
where the uncertainty corresponds to that of the (3,0) moment, and the value
employed for $V_{us}$ is the central PDG average
$|V_{us}|=0.2196 \pm 0.0026$ \cite{pdg:02}, based on the analyses
\cite{lr:84,cknrt:01,cl:02}. (See also ref.~\cite{CKMBook}.) To illustrate
the dependence of $m_s$ on $V_{us}$, one can compare to the result when using
the value $|V_{us}|=0.2225 \pm 0.0021$ obtained from a fit imposing unitarity
on the CKM matrix \cite{pdg:02}. The strange quark mass is then found to be
\begin{equation}
\label{msUT}
m_s(2\,\gev) = 117 \pm 17 \, \mev \,.
\end{equation}
The dominant theoretical uncertainties in the results of eqs.~\eqn{msLR} and
\eqn{msUT} originate from higher order perturbative corrections as well as
the SU(3) breaking ratio of the quark condensates
$\langle\bar ss\rangle/\langle\bar qq\rangle$ \cite{jam:02} which arises in
the next order dimension-4 contribution to eq.~\eqn{delRtaukl}. A detailed
discussion of all input parameters and a breakup of the various uncertainties
can be found in ref.~\cite{gjpps:03}.

Before coming to the determination of $V_{us}$ from the $\tau$ sum rule, a
somewhat disturbing observation in the strange mass analysis still needs to
be discussed. Namely, it is found that $m_s$ obtained from the lowest (0,0)
moment is largest, and then the strange mass continually decreases for the
higher moments. Although the individual $m_s$ values are all within one sigma
of the central result, it would be important to find out if this $k$-dependence
stems from a deficiency in the theoretical description or in the experimental
data.

One possible explanation of the effect could be missing contributions in the
higher-energy region of the experimental strange spectrum, because these
would suppress the strange mass determined from lower-$k$ moments. A hint in
this direction is given by a recent CLEO result \cite{cleo:03}, confirming
previous CLEO and OPAL results, which found the branching fraction of the
exclusive $K\pi\pi$ mode to be
$B(\tau^-\to K^-\pi^+\pi^-\nu_\tau)=(3.84\pm 0.14\pm 0.38)\cdot 10^{-3}$,
nearly 3 sigma higher than the corresponding findings by ALEPH \cite{aleph:99},
on which the spectral moment analysis is based.

\section{A novel route to \boldmath{$V_{us}$}}

\begin{figure}[thb]
\begin{center}
\includegraphics[angle=270, width=7cm]{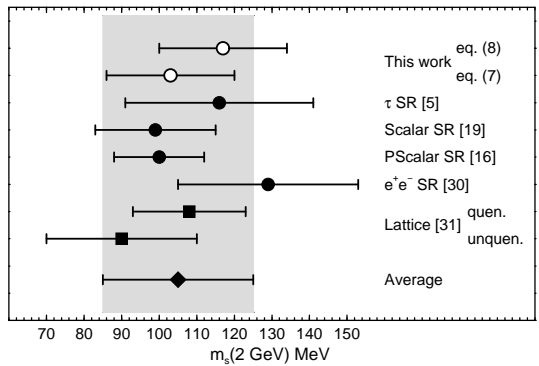}
\end{center}
\caption{Summary of recent results for $m_s(2\,\gev)$.
\label{fig:ms}}
\end{figure}

In order to calculate $V_{us}$ from the SU(3)-breaking difference
\eqn{delRtaukl}, we now require a value for the strange mass from other
sources as an input so that we are in a position to calculate
$\delta R_\tau^{kl}$ from theory. In the following, we shall use the
result $m_s(2\,\gev)=105\pm 20\,\mev$, a value compatible with most recent
determinations of $m_s$ from QCD sum rules
\cite{gjpps:03,cdghpp:01,jop:02,mk:02,nar:99} and lattice QCD
\cite{wit:02} A compilation of recent strange mass determinations is also
displayed in figure \ref{fig:ms}.

Since the sensitivity to $V_{us}$ is strongest for the (0,0) moment, where
also the theoretical uncertainties are smallest, this moment will be used
for the determination of $V_{us}$. Inserting the above strange mass value
into the theoretical expression for $\delta R_{\tau,th}$, one finds
\cite{gjpps:03}
\begin{equation}
\label{delR00th}
\delta R_{\tau,th} \;=\; 0.229 \pm 0.030 \,,
\end{equation}
where the uncertainty dominantly results from a variation of the strange quark
mass within its errors. Assuming unitarity of the CKM matrix in order to
express the CKM matrix element $V_{ud}$ in terms of $V_{us}$ in
eq.~\eqn{delRtaukl}, we then obtain
\begin{equation}
\label{VusUT}
V_{us} = 0.2179 \pm 0.0044_{\rm exp} \pm 0.0009_{\rm th} =
0.2179 \pm 0.0045\,.
\end{equation}
The first given error is the experimental uncertainty due to
$R_{\tau}=3.642\pm 0.012$ and, most importantly, the hadronic $\tau$ decay
rate with net strangeness $R_{\tau,S}=0.1625\pm 0.0066$ \cite{dav:02}, whereas
the second error stems from the theoretical value for $\delta R_{\tau}$. Even
though the theoretical error on $\delta R_{\tau,th}$ is roughly 15\%, since 
individually both terms on the rhs of \eqn{delRtaukl} are much larger, for the
extraction of $V_{us}$, $\delta R_{\tau,th}$ is only a correction and its
error rather unimportant. The theoretical uncertainty in $\delta R_{\tau,th}$
will only start to matter once the experimental error on $R_{\tau,S}$ is much
improved, possibly through an  analysis of the BABAR and BELLE $\tau$ data
samples.

\section{Conclusions}

Taking advantage of the strong sensitivity of the flavour-breaking $\tau$
sum rule on the CKM matrix element $V_{us}$, it is possible to determine it
from hadronic $\tau$ decay data. This requires a value of the strange quark
mass as an input which can be obtained from other sources like QCD sum rules
or the lattice. The result for $V_{us}$ thus obtained is
\begin{equation}
V_{us} \;=\; 0.2179 \pm 0.0045\,,
\end{equation}
where the error is largely dominated by the experimental uncertainty on
$R_{\tau,S}$. A reduction of this uncertainty by a factor of two would result
in a corresponding reduction of the error on $V_{us}$, which would lead to a
determination more precise than the current PDG average.

Furthermore, precise experimental measurements of $R_{\tau,S}^{kl}$ and the
SU(3)-breaking differences $\delta R_\tau^{kl}$ would open the possibility to
determine both $m_s$ and $V_{us}$ simultaneously. This can hopefully be
achieved with the BABAR and BELLE $\tau$ data samples in the near future.
This is particularly important since a very recent new measurement of $K_{e3}$
decays \cite{e865:03} points to a larger value of $V_{us}$, almost 3 sigma
away from the present PDG average.

\medskip
\subsection*{Acknowledgements}
It is a pleasure to thank A.~Pich for helpful discussions. This work was
supported by the {\em Deutsche Forschungsgemeinschaft} through a Heisenberg
fellowship.

\end{document}